\documentclass[12pt]{article}
\usepackage{amsmath,latexsym}
\usepackage{graphicx}
\begin{document}

 \title{\Huge A Theoretical Construction of Thin Shell Wormhole from Tidal Charged Black hole }
 \author{F.Rahaman$^*$ $^{\dag}$, M.Kalam$^{\ddag}$ , K A Rahman$^*$  and  S.Chakraborti$^*$  }
\date{}
 \maketitle

 \begin{abstract}
Recently, Dadhich et al [ Phys.Lett.B 487, 1 (2000)] have
discovered a black hole solution localized on a three brane in
five dimensional gravity in the Randall-Sundrum scenario. In this
article, we develop a new class of thin shell wormhole by
surgically grafting above two black hole spacetimes. Various
aspects of this thin wormhole are also analyzed.
\end{abstract}

  \footnotetext{ Pacs Nos :  04.20 Gz,04.50 + h, 04.20 Jb

 Key words:  Thin shell wormhole, Brane worlds,   Stability

 $*$Dept.of Mathematics, Jadavpur University, Kolkata-700 032,
 India:

 E-Mail:farook\_rahaman@yahoo.com\\
 $^{\dag}$  Centre for Theoretical
          Physics,
          Jamia Millia  Islamia,  New Delhi  - 110025

$\ddag$Dept. of Phys. , Netaji Nagar College for Women, Regent Estate, Kolkata-700092, India.\\
    \mbox{} \hspace{.2in}}

\title{\Huge Introduction: }

In recent years, several scientists around the world have given
their attention to the brane world gravity. Within the framework
of brane world, one can explain one of the hierarchy problems in
the current standard models of high energy physics [1]. Also,
brane world framework may give clue how to solve the greatest
challenging problem in theoretical physics namely the unification
of all fundamental forces in nature [2]. Actually, the string
theory was the inspiration behind the idea of brane world
scenario. The brane world idea is that the matter fields are
located on a three dimensional subspace, called brane embedded in
1 + 3 + d dimensions in which the gravity can propagate in the
d-extra dimensions. Here, the d-extra dimensions need not all be
small or even compact. Most of the recent studies consider a
simple version of the brane world scenario where all matters (
except gravity ) are confined to a 3-brane embedded in a five
dimensional spacetime (bulk) while gravity can propagate in the
bulk.

\pagebreak

 As a consequence, the gravity on the brane can be
described by the modified 4-dimensional Einstein equations which
contains (i) $ S_\mu^\nu$ , which is quardratic in the stress
energy tensor of matter confined on the brane (ii) the trace less
tensor  $ E_\mu^\nu$, originating from the 5D Weyl tensor and
describing tidal effects on the brane from the bulk geometry.
Recently, Dadhich et al [3] have presented a spherically symmetric
solution which describes a black hole localized on a three brane
in five dimensional gravity in the brane world scenario. This
black hole ( without electric charge) is termed as tidal charged
black hole. In this case tidal charge is arising via gravitational
effects from the fifth dimension.

Motivated by Morris and Thorne's work [4], the study of
traversable wormhole have been a focus of  interest in recent
years. These are the solutions of Einstein's equations that have
two regions connected by a throat. To get a wormhole solution,
one has to tolerate the violation of null energy condition. In
other words, the presence of exotic matter ( i.e. the matter
which violates the null energy condition ) is unavoidable to get
a wormhole solution. As it is difficult to deal with exotic
matter, it is useful to minimize the usage of exotic matter. In
recent past, Visser [5]  has proposed a way, which is known as
'Cut and Paste' technique,  of minimizing the usage of exotic
matter to construct a wormhole in which the exotic matter is
concentrated at the wormhole throat. In 'Cut and Paste' technique,
the wormholes are theoretically constructed by cutting and
pasting two manifolds to obtain geodesically complete new
manifold with a throat placed in the joining shell. Using
Darmois-Israel [6] formalism, one can determine the surface
stresses of the exotic matter ( located in thin shell placed at
the joining surface ). Recently, several physicists are interested
to develop thin shell wormholes.
 Visser and
Poisson  have  analyzed the stability of thin shell wormhole
constructed by joining the two Schwarzschild geometries [7]. The
stability of transparent spherical symmetric thin-shell wormholes
was examined by Ishak and Lake [8] . Eiroa and Romero [9] have
studied the linear stability of charged thin shell wormholes
constructed by joining the two Reissner-Nordstr\"{o}m spacetimes
under spherically symmetric perturbations.  Lobo and Crawford [10]
have extended the linear stability analysis to the thin shell
wormholes with Cosmological Constant.  Eiroa and Simeone [11] have
studied cylindrically symmetric thin shell wormhole geometries
associated to gauge cosmic strings. Also, the same authors have
constructed a charged thin shell wormhole in dilaton gravity and
they have shown that the reduction of the total amount of exotic
matter is dependent on the Dilaton-Maxwell coupling parameter
[12]. The five dimensional thin shell wormholes in
Einstein-Maxwell theory with a Gauss Bonnet term has been studied
by Thibeault et al [13]. They have made a linearized stability
analysis under radial perturbations. Recently, the present authors
have studied thin shell wormholes in higher dimensional
Einstein-Maxwell theory which is constructed by Cutting and
Pasting two metrics corresponding to a higher dimensional
Reissner-Nordstr\"{o}m black hole [14]. Also, the same authors
have constructed  thin shell wormhole in heterotic string theory
[15]. In this article, we study thin shell wormhole in Brane
worlds. We develop the model by cutting and pasting two metrics
corresponding to a tidal charged black hole. According to Dadhich
et al [3] tidal force arising via gravitational effects from the
fifth dimension i.e. it is arising from the projection onto the
brane of free gravitational field effects in the bulk.

The tidal charge 'q' of Dadhich et al's brane world black hole
contains the information of the extra dimension and does affect on
the geodesics as well as on the gravitational potential. So,  it
is of great interest to investigate how the tidal force affects
on the thin shell wormhole. Various aspects of these thin shell
wormhole, namely, temporal evolution of the throat, stability,
total amount of exotic matter will be discussed. We have shown
the tidal charge affects significantly on stability of this thin
shell wormhole. It is also shown that the total amount of exotic
matter is reduced by increasing of the tidal charge.

The layout of the paper as follows :

In section 2, the reader is reminded about tidal charged black
hole obtained by Dadhich et al. In section 3, thin shell wormhole
has been constructed by means of the Cut and Paste techniques. In
section 4, the time evolution of the radius of the throat is
considered whereas  linearized stability analysis is studied in
section 5.  Section 6 is devoted to a brief summary and discussion
including the calculation of the total amount of exotic matter
needed.

\title{\Huge2. Brane world Black holes : }

The gravitational field equations induced on the brane is
described by modified Einstein equations from 5-dimensional
gravity with the aid of Gauss-Codazzi equations to the effective
4-dimensional field containing the new terms carrying bulk
effects onto the brane as [1-2]
\begin{equation}
              G_{\mu\nu} =  \Lambda g_{\mu\nu} + k_4^2 T_{\mu\nu} + k_5^4 S_{\mu\nu}
             - E_{\mu\nu}
           \end{equation}
           Here $\Lambda$ is the 4D cosmological constant
           expressed in terms of the 5D , ${\Lambda}_5$ and brane
           tension $ \lambda$:

           $ \Lambda = \frac{1}{2} ({\Lambda}_5  +
           \frac{1}{6}k_5^4 \lambda)$ ;

           $ k_4^2 = 8 \pi G  = \frac{k_5^4\lambda}{6\pi} $

           where $k_4$ is the 4D gravitational constant and G is
           the Newton's constant of gravity.
           $T_{\mu\nu}$ is the usual energy momentum tensor of
           matter on the brane and the local bulk effects on the
           matter, $ S_{\mu\nu}$ consists of squares of
           $T_{\mu\nu}$ ( it is a local higher energy correction
           term). $E_{\mu\nu}$ consists of the projection of the
           bulk Weyl tensor onto the brane ( it is non local from
           the brane point of view ). By construction, $E_\mu^\nu$
           is  trace less, $E_\mu^\mu = 0$ and on the brane, in
           the vacuum case, it satisfies conservation equation $
           \nabla^\mu E_\mu^\nu = 0 $. Inspite of the absence of
           matter, $E_\mu^\nu$ may not be zero.

\pagebreak

           Considering the static spherically symmetric spacetime,
\begin{equation}
               ds^2=  f(r) dt^2 - \frac{dr^2}{ f(r)} - r^2
               d\Omega_2^2,
               \end{equation}
and using the vacuum brane field equation ( with $\Lambda$ set
equal to zero)

$G_\mu^\nu = - E_\mu^\nu$,

Dadhich et al [3], have obtained a new black hole solution as

\begin{equation}
               ds^2=  (1 - \frac{2m}{r} + \frac{q}{r^2}) dt^2 - \frac{dr^2}{
               (1 - \frac{2m}{r} + \frac{q}{r^2})} - r^2
               d\Omega_2^2,
               \end{equation}
               These black holes are characterized by two
               parameters : their mass and dimension less tidal
               charge, q.  The tidal  charge  parameter
               q  comes  from  the projection    on the brane of
               free gravitational field   effects  in  the  bulk. Here the tidal charge q can take both positive
               and negative values. When the tidal charge takes
               positive values, then the metric (3) is analogous
               to Reissner-Nordstr\"{o}m Black hole solution. For
               $ q < m^2$, it describes tidal charge black hole
               with two horizons at $ r_h = m \pm \sqrt{m^2 - q}$,
               which are below the Schwarzschild radius. Now for $
               q < 0 $, the above tidal charge black hole has only
               one horizon at $ r_h = m + \sqrt{m^2 +
               \mid q \mid}$, above the Schwarzschild radius. The
               gravitational field of this black hole is
               increased due the presence of the tidal charge.

\title{\Huge3. Cut and Paste Tactics: }

Dadhich et al [3]  have given exact localized black hole solution
 in which the brane is located at $ \chi = 0 $ ( $ \chi$ is the
 fifth coordinate ). This black hole formed from collapsed matter
 confined on brane. The induced four dimensional metric on the
 brane is given in equation (3). We work with ( 3 + 1 )
 dimensional brane, which is contained in a ( 4 + 1 ) dimensional
 bulk and consider the 4-dimensional horizon structure of the
 brane world black hole because like Dadhich et al [3], we are not
 interested to take the effect of the  brane world black hole on
 the bulk geometry. And for that we do not consider the bulk
 metric.
 Thus, to construct thin shell wormhole in brane world, we cut two
copies of region from the tidal charged black hole geometry
described  by $ \Omega^\pm = ( x \mid r \leq a )  $, where $
a\geq r_h$ ( position of event horizon). In this study, we assume
the case, where tidal charge is negative i.e. horizon of this
black hole is greater than the Schwarzschild horizon. Now taking
the two copies of the remaining regions, $ M^\pm = ( x \mid r
\geq a )  $, we paste the two pieces together at the hypersurface
$ \Sigma = \Sigma^\pm = ( x \mid r = a )  $. This surgical
grafting  produces a geodesically complete manifold $ M = M^+
\bigcup M^- $ with a matter shell at the surface $ r = a $ ,
where the throat of the wormhole is located. Thus a single
manifold M is obtained which connects two asymptotically flat
regions at their boundaries $\Sigma$ and the throat is placed at
$\Sigma$ ( here $\Sigma$ is a synchronous time like hypersurface
). Since M is a piece wise tidal charged black hole spacetime, the
stress energy tensor is either every where zero or obeys all
energy conditions except at the throat itself.

At $\sum$, one expects that the stress energy tensor to be
proportional to a delta function. Following Darmois-Israel
formalism, we shall determine the surface stresses at the junction
boundary. The intrinsic coordinates in $\Sigma$ are taken as $
\xi^i = ( \tau, {\theta}, \phi)$  with $\tau$ is the proper time
on the shell.

To understand the dynamics of the wormhole, we assume  the radius
of the throat be a function of the proper time $ a = a(\tau)$. The
parametric equation for $\Sigma$ is defined   by
\begin{equation}\Sigma : F(r,\tau ) = r - a(\tau)\end{equation}
The extrinsic curvature associated with the two sides of the
shell are
\begin{equation}K_{ij}^\pm =  - n_\nu^\pm\ [ \frac{\partial^2X_\nu}
{\partial \xi^i\partial \xi^j } +
 \Gamma_{\alpha\beta}^\nu \frac{\partial X^\alpha}{\partial \xi^i}
 \frac{\partial X^\beta}{\partial \xi^j }] |_\Sigma \end{equation}
where $ n_\nu^\pm\ $ are the unit normals to $\Sigma$,
\begin{equation} n_\nu^\pm =  \pm   | g^{\alpha\beta}\frac{\partial F}{\partial X^\alpha}
 \frac{\partial F}{\partial X^\beta} |^{-\frac{1}{2}} \frac{\partial F}{\partial X^\nu} \end{equation}
with $ n^\mu n_\mu = 1 $.

The intrinsic metric on $\Sigma$ is given by

\begin{equation}
               ds^2 =  - d\tau^2 + a(\tau)^2 d\Omega_2^2
               \end{equation}

 From Lanczos equation, one can obtain the surface stress
energy tensor $ S_j^i = diag ( - \sigma , -v_{\theta}, -v_{\phi})
$ (where $ \sigma$ is the  surface  energy density and $
v_{\theta,\phi} $ , the surface tensions)  as
\begin{equation}
               \sigma =  - \frac{1}{2\pi a}\sqrt{1-\frac{2m}{a} - \frac{Q}{a^2} + \dot{a}^2}
               \end{equation}

\begin{equation}
              - v_{\theta} = - v_{\phi}
 = - v  =  \frac{1}{4\pi a}\frac{1 - \frac{m}{a} + \dot{a}^2 + a \ddot{a} }{\sqrt{1-\frac{2m}{a}
  - \frac{Q}{a^2} + \dot{a}^2}}
               \end{equation}

where over dot  means the derivative with respect to $\tau$ and $
q = - Q, Q > 0 $.

Negative surface energy density in (8) implies the existence of
exotic matter at the shell. The negative signs of the tensions
mean that they are indeed pressures.

\pagebreak

\title{\Huge4. Time evolution of radius of the throat: }

The static equations are obtained with $\dot{a} = 0 $ and
$\ddot{a}= 0 $ in equations (8) and (9):
\begin{equation}
               \sigma_0 =  - \frac{1}{2\pi a}\sqrt{1-\frac{2m}{a} -
               \frac{Q}{a^2}}
               \end{equation}
\begin{equation}
  v_0  = -  \frac{1}{4\pi a}\frac{1 - \frac{m}{a}}{\sqrt{1-\frac{2m}{a}
  - \frac{Q}{a^2}}} \end{equation}

[ suffix 0 indicates the static situation ]

  Now, one can write the equations (10) and (11)in the form
\begin{equation}
              v_0 =  w(a)  \sigma_0
               \end{equation}
where
\begin{equation}
              w(a) = \frac{1}{2}\frac{1 - \frac{m}{a}}{(1-\frac{2m}{a}
  - \frac{Q}{a^2})}
               \end{equation}
Following Eiroa et al and C Bejarano et al [12], we assume that
the equation of state does not depend on the derivative of
$a(\tau)$ i.e. it is the same form as in the static one. Now
putting $\sigma $, v in place of $\sigma_0$ , $v_0$ from (8) and
(9) in (12), we get the following expression as
\begin{equation}
             \ddot{a}( a - 2m
  - \frac{Q}{a})  -  \dot{a}^2 ( \frac{m}{a} + \frac{Q}{a^2})= 0
               \end{equation}
This implies,
\begin{equation}
             \dot{a}(\tau) =   \dot{a}(\tau_0)[ \frac { 1 - \frac{2m}{a(\tau)}
             -
             \frac{Q}{a^2(\tau)}}{1 - \frac{2m}{a(\tau_0)} -
             \frac{Q}{a^2(\tau_0)}}]^{\frac{1}{2}}
               \end{equation}
               Here, $\tau_0$ is arbitrary fixed time.

               Thus we get,

            $ \int_{a(\tau_0)}^{a(\tau)}
               \frac{da}{[1 - \frac{2m}{a(\tau)} -
             \frac{Q}{a^2(\tau)}]^{\frac{1}{2}}} =
             \dot{a}(\tau_0)( \tau - \tau_0) [1 - \frac{2m}{a(\tau_0)}
             - \frac{Q}{a^2(\tau_0)}]^{-\frac{1}{2}}
               $

This gives,
   \begin{equation} [a^2(\tau) - 2m a(\tau) - Q ]^{\frac{1}{2}} + m
 \ln [ 2(a^2(\tau) - 2m a(\tau) - Q )^{\frac{1}{2}} + 2a(\tau)
 -
 2m ] = \frac{\dot{a}(\tau_0)( \tau - \tau_0)}{(1 - \frac{2m}{a(\tau_0)}
             \frac{Q}{a^2(\tau_0)})^{\frac{1}{2}}}
                \end{equation}

 The above implicit expression gives the time evolution
of the radius of the throat.

\pagebreak

 The velocity and acceleration of the throat are
\begin{equation}
             \dot{a}(\tau) =   \dot{a}(\tau_0)[ \frac { 1 - \frac{2m}{a(\tau)}
             -
             \frac{Q}{a^2(\tau)}}{1 - \frac{2m}{a(\tau_0)} -
             \frac{Q}{a^2(\tau_0)}}]^{\frac{1}{2}}
               \end{equation}
               and
\begin{equation}
             \ddot{a}(\tau) = \dot{a}^2(\tau_0)[ \frac { \frac{m}{a^2(\tau)}
             +
             \frac{Q}{a^3(\tau)}}{1 - \frac{2m}{a(\tau_0)} -
             \frac{Q}{a^2(\tau_0)}}]
                 \end{equation}

From the above two expressions, one can easily see that the sign
of the velocity is given by the sign of the initial velocity and
the acceleration is always positive. It is immaterial whether the
initial velocity is positive or negative, the throat expands
forever. This would imply that the equilibrium position is always
unstable. However, if the initial velocity is zero, the velocity
and acceleration of the throat would be zero i.e. throat be in
static equilibrium position. Now, we shall study the stability of
the configuration under small perturbations around static solution
situated at $a_0$ ( initial velocity will be assumed to be zero ).

\title{\Huge5. Linearized Stability Analysis: }

Rearranging equation (8), we obtain the thin shell's  equation of
motion

            \begin{equation}  \dot{a}^2 + V(a)= 0  \end{equation}

                Here  the potential is defined  as
\begin{equation}
              V(a) =  1-\frac{2m}{a} - \frac{Q}{a^2} - 4\pi^2 a^2\sigma^2(a)
                 \end{equation}

 Linearizing around a static solution situated at $a_0$,
one can expand V(a) around $a_0$ to yield
\begin{equation}
              V =  V(a_0) + V^\prime(a_0) ( a - a_0) + \frac{1}{2} V^{\prime\prime}(a_0)
              ( a - a_0)^2 + 0[( a - a_0)^3]
                 \end{equation}
where prime denotes derivative with respect to $a$.

Since we are linearizing around a static solution at $ a = a_0 $,
we have $ V(a_0) = 0 $ and $ V^\prime(a_0)= 0 $. The stable
equilibrium configurations correspond to the condition $
V^{\prime\prime}(a_0)> 0 $. Now we define a parameter $\beta$,
which is interpreted as the speed of sound, by the relation [7]
\begin{equation}
              \beta^2(\sigma) = \frac{ \partial p}{\partial
              \sigma}|_\sigma
                 \end{equation}
Here,
\begin{equation} V^{\prime\prime}(a) = -\frac{4m}{a^3} - \frac{6Q}{a^4} - 8\pi^2 \sigma^2
  - 32\pi^2 a \sigma \sigma^\prime   - 8\pi^2 a^2 (\sigma^{\prime})^2
 - 8\pi^2 a^2\sigma \sigma^{\prime\prime}
 \end{equation}

 Since the negative tension is equivalent to pressure, we take $
 -v = -v_\theta = -v_\phi = p $.
 From equations (8) and (9), one can write energy conservation
 equation as
\begin{equation}
               \dot{\sigma} +  2\frac{\dot{a}}{a}( p + \sigma ) = 0
               \end{equation}
               or
 \begin{equation}
               \frac {d}{d \tau} (4\pi\sigma a^2) + p \frac{d}{d \tau}(4\pi a^2)= 0
               \end{equation}

From equation (24) ( by using (22) ),  we obtain,

\begin{equation}
               \sigma^{\prime\prime} +  \frac{2}{a} \sigma ^{\prime} ( 1 + \beta^2)
               - \frac{2}{a^2} ( p + \sigma) = 0
               \end{equation}

Now, the second derivative of the potential is taken the
following form as
\begin{equation}
              V^{\prime\prime}(a_0) =  - \frac{ 2}{a_0^4}[  a_0[ \frac {(a_0-m)^3 + m(m^2+Q)}{a_0^2 -2ma_0 - Q}] + 2( a_0^2
              -3ma_0 -2Q)\beta_0^2)]
                 \end{equation}

 The wormhole solution is stable if $
V^{\prime\prime}(a_0)> 0 $ i.e. if
\begin{equation} ( a_0^2
              -3ma_0 -2Q)\beta_0^2 < -a_0[ \frac {(a_0-m)^3 + m(m^2+Q)}{2(a_0^2 -2ma_0 - Q})]  \end{equation}
The right hand side of this inequality is negative because, $ a_0
> m( 1 + \sqrt{1+\frac{Q}{m^2}}) \equiv r_h $. The left hand side
flips sign at $ a_0 = \frac{3}{2}m[
1+\sqrt{1+\frac{8Q}{9m^2}}\equiv r_+ > r_h$. Therefore, if one
treats $a_0$, m and Q   are specified quantities, then the
stability of the configuration requires the following restrictions
on the parameter $\beta_0$.

$ \beta_0^2 < -a_0[ \frac {(a_0-m)^3 + m(m^2+Q)}{2(a_0^2 -2ma_0 -
Q)( a_0^2 -3ma_0 -2Q)}] $     if,  $ a_0 > r_+ $

$ \beta_0^2 > -a_0[ \frac {(a_0-m)^3 + m(m^2+Q)}{2(a_0^2 -2ma_0 -
Q)( a_0^2 -3ma_0 -2Q)}] $     if,  $ r_h < a_0 <  r_+ $

This means there exists some part of the parameter space where the
throat location is stable. For a lot of useful information, we
show the stability regions graphically. For normal situation i.e.
in case of real matter, $ \beta_0$ represents  the velocity of
sound  and it lies within  the interval,   $ \beta^2_0$  $
\epsilon $ $ (0,1]$. However, in the  presence of exotic matter (
as it happens to be   in the throat ), $ \beta_0$  is not velocity
of sound. So for, exotic matter this range  may be relaxed. One
can see ref.[7] for   an extensive discussion on the respective
physical interpretation of $ \beta_0$ in the presence of exotic
matter.
\begin{figure}[htbp]
    \centering
        \includegraphics[scale=.8]{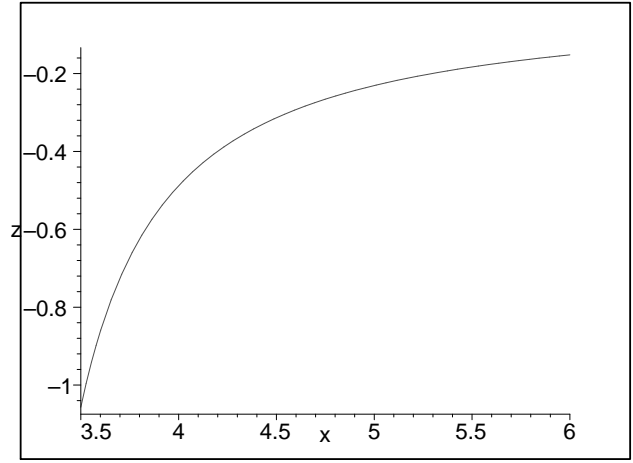}
        \caption{ Here we plot $ z = \beta^2_{|{(a=a_0)}} $ $ Vs.$ $ x= \frac{a_0}{m}$ for  $ a_0 > r_+ $ ( $  \frac{Q}{m} = .1 $)  .
         The stability region is situated below the curve.}
    \label{fig:stability}
\end{figure}

\begin{figure}[htbp]
    \centering
        \includegraphics[scale=.8]{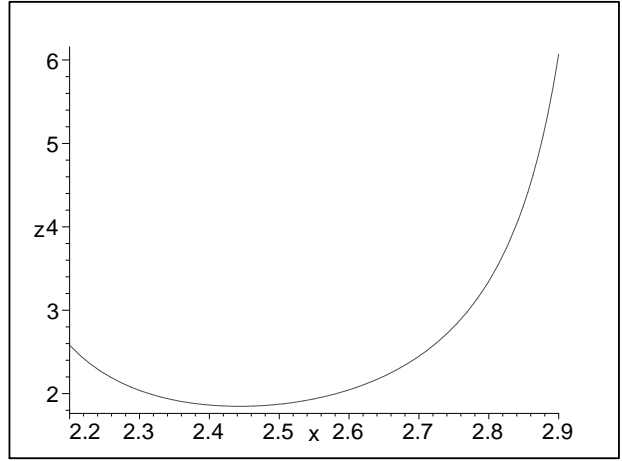}
        \caption{ Here we plot $ z = \beta^2_{|{(a=a_0)}} $ $ Vs.$ $ x= \frac{m}{a_0}$  for  $ r_h < a_0 <  r_+ $ ($  \frac{Q}{m} = .1 $).
         The stability region is situated above the curve.}
    \label{fig:stability}
\end{figure}

\pagebreak

\title{\Huge6. Summary  and Discussions: }

In this article, we have studied thin shell wormhole in brane
world scenario by surgically grafting two tidal charged black
hole spacetimes. Though the tidal charge parameter 'q' of this
black hole can take both positive and negative values , but in
this study, we consider the case for $ q < 0 $ [ For $ q
> 0 $, the solution is analogous to the Reissner-Nordstr\"{o}m Black hole
solution ]. In this case, the black hole has only one horizon,
which lies outside the Schwarzschild horizon. We have considered
an equation of state that relates the tension with the surface
energy density of the exotic matter at the throat. We have
obtained the time evolution of the radius of throat. One could see
that whether the initial velocity is positive or negative, the
throat expands indefinitely. But, when initial velocity is zero,
then the radius of the throat remains constant i.e. the throat be
in a static equilibrium position. We have analyzed the dynamical
stability of the thin shell, considering linearized radial
perturbations around static solution. To analyze this, we define a
parameter $ \beta^2 = \frac{p^\prime }{\sigma^\prime} $ as a
parametrization of the stability of equilibrium. We have obtained
a restriction on $\beta^2$ to get stable equilibrium of the thin
wormhole( see eq.(28)).

 The total amount of exotic matter for the thin wormhole can be
quantified by the integral ( In this case, radial pressure, $ p_r
= 0 $ and we have $ \rho  < 0 ,  \rho + p_r <0 $ i.e. both energy
conditions are violated. The transverse pressure is $ p_t =
p_\theta = p_\phi = - v $ and one can see from (10) and (11)  that
the sign of $ \rho + p_t $ is not fixed but depends on the value
of the parameters )

$
             \Omega =  \int [\rho + p_r]
\sqrt{-g}d^3x
                $ .

 Following Eiroa and Simone [11] , we introduce a
new radial coordinate $ R  =  \pm ( r -a ) $ in M ( $\pm $ for
$M^{\pm}$ respectively ) so that

$
            \Omega =  \int_0^{2\pi} \int_0^{\pi}\int_{-\infty}^\infty [\rho + p_r]
\sqrt{-g}dRd{\theta}d{\phi}
                 $

Since the shell does not exert radial pressure and the energy
density is located on a thin shell surface, so that $ \rho =
\delta(R)\sigma_0$, then we have

$             \Omega = \int_0^{2\pi} \int_0^\pi  [\rho \sqrt{-g}
]|_{r=a_0} d{\theta}d{\phi} = 4\pi a_0^2 \sigma(a_0)$

Thus one gets, $
             \Omega =  - 2a_0 \sqrt{1 -\frac{2m}{a_0} -
             \frac{Q}{a_0^2}}$  .

One could see that the tidal charge and mass of the black hole
affect the total amount of exotic matter needed. The variation of
the total amount of exotic matter with respect to tidal charge and
mass of the black hole is shown in the figure 3.

\begin{figure}[htbp]
    \centering
        \includegraphics[scale=.73]{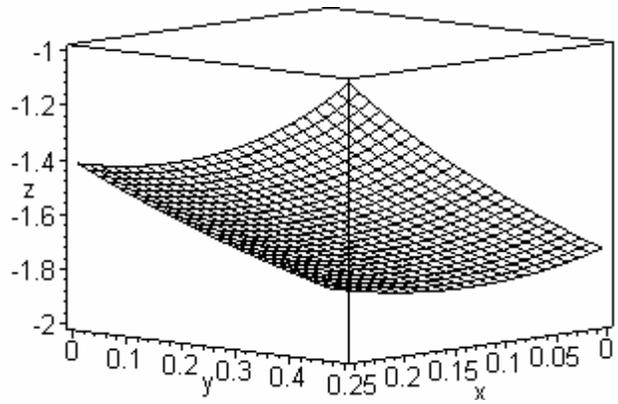}
        \caption{ Here we plot $ z = \frac{\Omega }{a_0} $ $ Vs.$ $ x= \frac{m}{a_0}$ and $ y = \sqrt{\frac{Q}{a_0^2}} $  .
        }
    \label{fig:newbrane5}
\end{figure}

\pagebreak

Now we are interested to the fact that under what conditions the
total amount of exotic matter could be reduced. If mass of the
black hole remains fixed , then the total amount of exotic matter
is reduced by increasing the tidal charge. Thus tidal charge
contains the information of extra dimension plays significant
role to reducing total amount of exotic matter needed.

Also if tidal charge of the black hole is kept fixed, then the
total amount of exotic matter is reduced by increasing the mass of
the black hole. Thus one can see that less exotic matter is needed
when tidal charge and mass of the black hole are increased. These
are depicted in figure 4 and 5.

\begin{figure}[htbp]
    \centering
        \includegraphics[scale=.7]{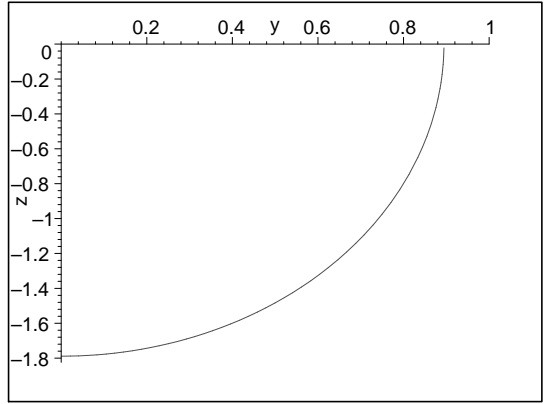}
        \caption{ Here we plot $ z = \frac{\Omega }{a_0} $  Vs. $ y = \sqrt{\frac{Q}{a_0^2}} $ for $ \frac{m}{a_0} = \frac{1}{10}$   .
        }
    \label{fig:newbrane2}
\end{figure}\begin{figure}[htbp]
    \centering
        \includegraphics[scale=.7]{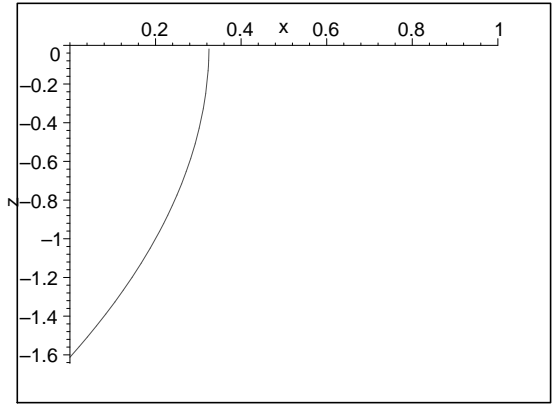}
        \caption{ Here we plot $ z = \frac{\Omega }{a_0} $ $ Vs.$ $ x= \frac{m}{a_0}$  for $ \sqrt{\frac{Q}{a_0^2}} = .35 $  .
        }
    \label{fig:newbrane3}
\end{figure}
             Further, from the above expression, one can see that $ \Omega$ approaches
to zero when wormhole radius tends to the event horizon ( i.e.
when $ a_0 \rightarrow r_h  $ ). So one can get vanishing amount
of exotic matter by taking $a_0 $ near $r_h $. Thus, one can note
that the total amount of exotic matter needed to support
traversable wormhole can be made infinitesimal small by taking
wormhole radius near the event horizon of the tidal charged black
hole. This is depicted in the figure-6.

\begin{figure}[htbp]
    \centering
        \includegraphics[scale=.8]{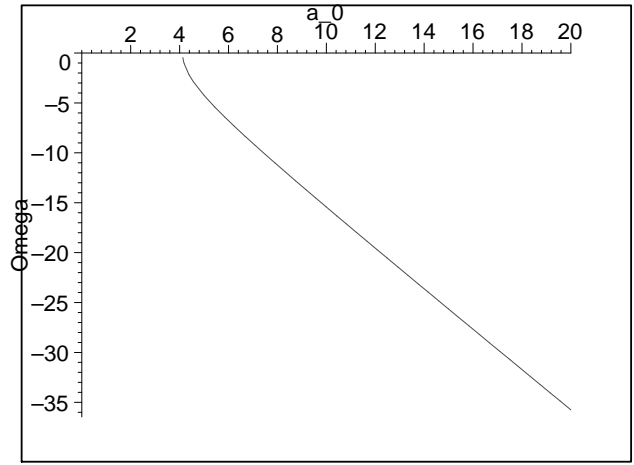}
        \caption{ We choose $ m  = 2 $ and
        $  Q = .5 $. The variation of total amount of exotic matter on the shell
        with respect to $a_0$ is shown in the figure.}
    \label{fig:brane2}
\end{figure}

\pagebreak

 {  \Huge Acknowledgments }

          F.R. is thankful to Jadavpur University and DST , Government of India for providing
          financial support. MK has been partially supported by
          UGC,
          Government of India under MRP scheme. We  are thankful to Centre for Theoretical
          Physics,
          Jamia Millia  Islamia  for worm  hospitality,
          where  a part  of the work has been carried out.
          Finally,  we are  grateful to  the referees for
          pointing out the errors and
          their constructive suggestions.
          \\


\end{document}